\documentclass[letterpaper]{article} 
\usepackage{aaai2026}  
\usepackage{times}  
\usepackage{helvet}  
\usepackage{courier}  
\usepackage[hyphens]{url}  
\usepackage{graphicx} 
\usepackage{natbib}  
\usepackage{caption} 
\usepackage{algorithm}
\usepackage{algorithmic}
\usepackage{amsmath}
\usepackage{booktabs} 
\usepackage{enumitem} 
\usepackage{amsfonts}
\usepackage{csquotes}
\usepackage{booktabs}
\usepackage{graphicx} 
\usepackage{multirow}
\usepackage{graphicx}
\usepackage{subcaption}
\usepackage{placeins}
\usepackage[table]{xcolor}
\usepackage{pifont}
\usepackage{tcolorbox}

\usepackage{newfloat}
\usepackage{listings}
\DeclareCaptionStyle{ruled}{labelfont=normalfont,labelsep=colon,strut=off} 
\floatstyle{ruled}
\newfloat{listing}{tb}{lst}{}
\floatname{listing}{Listing}
\title{Personalize Before Retrieve: LLM-based Personalized Query Expansion for User-Centric Retrieval}

\author {
    Yingyi Zhang\textsuperscript{\rm {1,2,}}\thanks{Equal contribution.},
    Pengyue Jia\textsuperscript{\rm {2,}}\footnotemark[1],
    Derong Xu\textsuperscript{\rm {2,3}},
    Yi Wen\textsuperscript{\rm {2}},
    Xianneng Li\textsuperscript{\rm {1,}}\thanks{Corresponding author.},
    Yichao Wang\textsuperscript{\rm 4,}\footnotemark[2],
    Wenlin Zhang\textsuperscript{\rm {2}},
    Xiaopeng Li\textsuperscript{\rm {2}},
    Weinan Gan\textsuperscript{\rm 4},
    Huifeng Guo\textsuperscript{\rm 4},
    Yong Liu\textsuperscript{\rm 4},
    Xiangyu Zhao\textsuperscript{\rm {2,}}\footnotemark[2]
}
\affiliations {
    \textsuperscript{\rm 1} Dalian University of Technology, 
    \textsuperscript{\rm 2} City University of Hong Kong, \\
    \textsuperscript{\rm 3} University of Science and Technology of China,    \textsuperscript{\rm 4}Huawei Technologies Ltd.\\
    xianneng@dlut.edu.cn, wangyichao5@huawei.com, xianzhao@cityu.edu.hk
}

\usepackage{bibentry}

\begin{document}

\maketitle

\begin{abstract}
Retrieval-Augmented Generation (RAG) critically depends on effective query expansion to retrieve relevant information. However, existing expansion methods adopt uniform strategies that overlook user-specific semantics, ignoring individual expression styles, preferences, and historical context. In practice, identical queries in text can express vastly different intentions across users. 
This representational rigidity limits the ability of current RAG systems to generalize effectively in personalized settings. Specifically, we identify two core challenges for personalization: 1) user expression styles are inherently diverse, making it difficult for standard expansions to preserve personalized intent. 2) user corpora induce heterogeneous semantic structures—varying in topical focus and lexical organization—which hinders the effective anchoring of expanded queries within the user’s corpora space.
To address these challenges, we propose \textit{Personalize Before Retrieve} (\textbf{PBR}), a framework that incorporates user-specific signals into query expansion prior to retrieval.
PBR consists of two components: \textbf{P-PRF}, which generates stylistically aligned pseudo feedback using user history for simulating user expression style, and \textbf{P-Anchor}, which performs graph-based structure alignment over user corpora to capture its structure. Together, they produce personalized query representations tailored for retrieval.
Experiments on two personalized benchmarks show that PBR consistently outperforms strong baselines, with up to 10\% gains on PersonaBench across retrievers.
Our findings demonstrate the value of modeling personalization \emph{before} retrieval to close the semantic gap in user-adaptive RAG systems. 

\end{abstract}

\begin{links}
    \link{Code}{https://github.com/Applied-Machine-Learning-Lab/PBR-code}
\end{links}

\section{Introduction}

\begin{figure}[h]
    \centering
    \includegraphics[width=0.95\columnwidth]{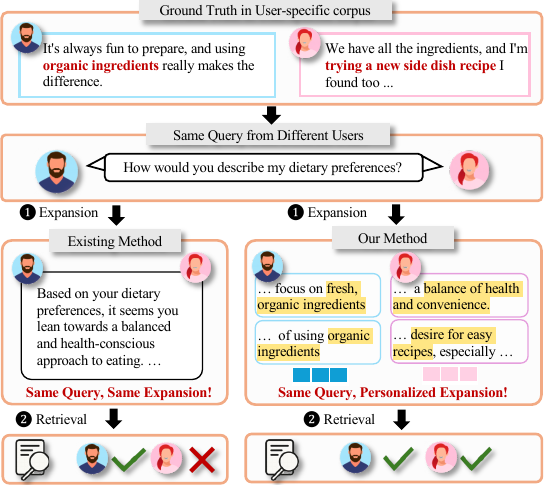} 
    \caption{An example comparing generic and personalized query expansion.}
    \label{fig:intro}
    \vspace{-20pt}
\end{figure}
Retrieval-Augmented Generation (RAG) has emerged as a pivotal paradigm for enhancing the capabilities of large language models (LLMs)~\cite{lewis2020retrieval, zhao2024retrieval,zhang2025lsrp}, leveraging a two-stage process: first retrieving relevant external corpus, then generating responses conditioned on the retrieved information~\cite{cheng2023lift}. 
The effectiveness of this paradigm hinges on the quality of query expansion by using LLM world knowledge~\cite{culpepper2021topic,song2024survey}, which directly impacts the accuracy and relevance of the retrieval stage. 
Most current strategies generate pseudo answers~\cite{gao2023precise} or candidate completions~\cite{jia2023mill} using LLMs in a zero-shot or few-shot manner. 
These generated semantic representations, often combined with the original query, are used to retrieve information from a global external corpus.

Recently, with LLMs widely adopted as personalized assistants, many user queries are required to retrieve from individual histories and contexts, demanding retrieval mechanisms that go beyond generic corpora to leverage user-specific data.
Existing query expansion strategies remain limited in their ability to capture user-specific semantics. 
In practice, the same textual query may convey vastly different intentions depending on a user’s preferences~\cite{zou2023users}, background knowledge~\cite{westhofen2025temporal}, or expression style~\cite{neelakanteswara2024rags}. As users exhibit diverse personas and contextual needs, it becomes increasingly important for query expansion methods to be personalized—capable of adapting to user-specific semantics within private corpora—so as to ensure accurate retrieval and enhance downstream generation.

As illustrated in Figure~\ref{fig:intro} (top), when two users submit the same query—``\emph{How would you describe my dietary preferences?}''—standard expansion methods yield generic outputs that fail to account for individual differences. This results in mismatched retrieval: for example, the health-conscious user (User~1) successfully retrieves content related to ``\emph{organic ingredients},'' which aligns with their preferences. In contrast, the variety-seeking user (User~2), whose personal content emphasizes ``\emph{trying new side dishes},'' fails to retrieve relevant information, highlighting the limitations of non-personalized expansion strategies. Such failures stem from neglecting rich signals in user—such as style, intent patterns, and thematic focus—resulting in semantically misaligned expansions and degraded retrieval performance.

Motivated by these limitations, we aim to develop a personalized query expansion framework that adapts to individual user intent and expression as shown in Figure~\ref{fig:intro}. However, introducing personalization into expansion brings two fundamental challenges:
\textbf{(1) User expression styles are inherently diverse.}  
Users articulate intent using varied linguistic patterns—ranging from minimal prompts to elaborative reasoning—driven by personal habits, domain familiarity, or rhetorical preference~\cite{neelakanteswara2024rags}. These styles are often implicit and non-transferable, making it difficult to construct expansions that faithfully preserve user-specific semantics within a generalizable framework.
\textbf{(2) User corpora induce heterogeneous semantic structures.} 
Users’ personal corpora often differ significantly in topical coverage, content organization, and linguistic granularity. This high degree of heterogeneity makes it difficult to locate reliable semantic regions related to the user query, and in turn, complicates the alignment of user-specific preferences. Without user-specific semantic anchors to guide expansion, queries are prone to drifting toward irrelevant regions, resulting in mismatched retrieval despite plausible surface semantics.
These challenges raise a central question: \textit{How can we personalize query expansion—both in style and structure—before retrieval to align with individual user intent and corpus characteristics?}

To address these challenges, we propose a \textit{Personalize Before Retrieve} (\textbf{PBR}) framework that integrates user-specific signals into query understanding prior to retrieval. PBR adopts a two-stage design that combines expression-level expansion with structure-level alignment, enabling queries to reflect both personal style and corpus semantics.
\textbf{\ding{182} P-PRF: Personal Style-Aligned Pseudo Relevance Feedback.}  
This module extracts linguistic patterns from user history to guide LLMs in generating pseudo-utterances and reasoning paths. These signals capture personalized expression style and hidden intent, often missed by generic expansion methods.
\textbf{\ding{183} P-Anchor: Personal Structure-Aligned Semantic Anchoring.}  
We encode user history into a semantic graph that captures localized relevance patterns. Personalized PageRank is applied to identify representative anchor points within this structure to encode the whole user corpora or to express user general preference. The query is then guided toward these anchors to reflect the user’s semantic landscape.
By jointly modeling stylistic variation and structural relevance, P-PRF and P-Anchor produce personalized query representations that enhance alignment between user intent and retrieval semantics in RAG systems. Our contributions can be summarized as follows:
\begin{itemize}[leftmargin=*]
    \item To the best of our knowledge, we are the first to propose a personalized query expansion framework for RAG systems that adapts to the user for improving retrieval.
    \item We propose \textbf{P-PRF}, a style-aware expansion module that generates pseudo queries and reasoning paths conditioned on user history.
    \item We introduce \textbf{P-Anchor}, a graph-based alignment module that grounds queries within personalized semantic spaces via local corpus structure.
    \item Experiments on two real-world dialogue datasets show that \textbf{PBR} significantly improves personalized retrieval, validating the benefits of pre-retrieval personalization.
\end{itemize}

\section{Problem Definition}

We study \textit{personalized query expansion}, which expands underspecified queries to align with both user expression style and corpus heterogeneity informed by past interactions.

Formally, let $q$ represent the user query, and let $H = \{h_1, h_2, \dots, h_n\}$ denote the user's historical utterances, referred to as the user corpus, where each $h_i$ is a textual segment from prior conversations or interactions. The query is encoded as $\mathbf{q} = \phi(q) \in \mathbb{R}^d$, while the historical utterances are encoded into a set of corpus vectors $\mathcal{C} = \{\mathbf{c}_1, \dots, \mathbf{c}_n\}$ using a fixed encoder $\phi(\cdot)$:
\begin{equation}
\mathbf{c}_i = \phi(h_i) \in \mathbb{R}^d, \quad \forall i = 1, \dots, n.
\label{eq:c_i}
\end{equation}

The goal is to construct a personalized query representation \(\mathbf{q}^* \in \mathbb{R}^d\) that retrieves content from the user corpus \(\mathcal{C}\) not only based on lexical-semantic similarity, but also aligned with user-specific latent intent. Formally, the retrieved set is defined as:
\begin{equation}
\mathcal{R}(\mathbf{q}^*) = \text{Top}_k\left( \left\{ \text{sim}(\mathbf{q}^*, \mathbf{c}_k) \mid \mathbf{c}_k \in \mathcal{C} \right\} \right),
\end{equation}
where \(\text{sim}(\cdot)\) denotes a similarity function (e.g., cosine similarity), and \(\text{Top}_k(\cdot)\) returns the top-\(k\) most relevant items.

The core challenge lies in constructing \(\mathbf{q}^*\) from the observable query \(\mathbf{q}\) and the latent user-specific context \(\mathcal{C}\). We formulate a transformation function \(f_\Theta(\mathbf{q}, \mathcal{C})\) such that:
\begin{equation}
\mathbf{q}^* = f_\Theta(\mathbf{q}, \mathcal{C}) = \mathbf{q} + \Delta_{\text{user}}(\mathbf{q}, \mathcal{C}),
\end{equation}
where \(\Delta_{\text{user}}(\mathbf{q}, \mathcal{C})\) encodes personalized adjustments reflecting individual expression style, intent formulation, and structural semantics in the user corpus.

\begin{figure*}[th]
    \centering
    \includegraphics[width=1.00\linewidth]{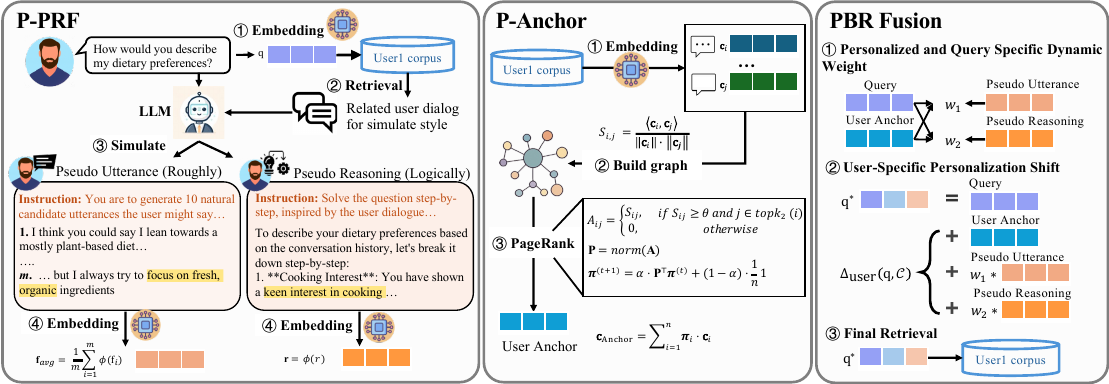}
    \caption{Overview of PBR, which consists of three components: \textbf{P-PRF} for stylistic expansion, \textbf{P-Anchor} for structural grounding, and \textbf{PBR Fusion} module for generating the final query representation. 
    }
    \vspace{-10pt}
    \label{fig:main_framework}
\end{figure*}

\section{PBR Framework}

\subsection{Framework Overview}
\label{sec:framework}

As illustrated in Figure~\ref{fig:main_framework}, the \textbf{PBR} framework enhances personalized retrieval by explicitly incorporating user-specific signals into query representation before retrieval. It consists of three core modules: \textbf{P-PRF} simulates two complementary forms of feedback to generate pseudo utterances and pseudo reasoning that reflect the user's stylistic tendencies and underlying reasoning logic. \textbf{P-Anchor} constructs a semantic graph over the user's corpus and applies PageRank to identify representative anchor points that capture the corpus-level user anchor. These two sources of personalized signals are integrated via \textbf{PBR Fusion}, which computes a personalized and query-specific dynamic weight to balance the contributions of pseudo utterances and pseudo reasoning. The resulting fused representation $q^*$ reflects both user hidden style and contextual grounding, enabling accurate retrieval from the semantic index constructed over the user corpus $C$.

\subsection{P-PRF}

Generic query expansion methods often overlook the nuanced expression styles~\cite{neelakanteswara2024rags} and implicit reasoning patterns behind user expression~\cite{li2025survey}, leading to mismatches in personalized retrieval. This challenge is exacerbated when queries are ambiguous or under-specified, requiring systems to infer intent beyond surface forms. Existing solutions typically apply uniform expansion strategies, failing to adapt to user-specific language signals.

To address these limitations, \textbf{P-PRF} simulates user-specific signals before retrieval by simulating personalized feedback in two complementary forms: \emph{roughly} via stylistic utterances, and \emph{logically} via intent reasoning. Rather than using the full user history $H$, we retrieve a task-relevant subset $H_q$  based on semantic similarity:
\begin{equation}
H_q= \text{TopK}\big(\{\text{sim}(\mathbf{q}, \mathbf{c}_t)\}_{t=1}^n\big), \quad |H_q| = k_1,
\end{equation}
where $\mathbf{q}$ and $\mathbf{c}_t$ are the embeddings of the query and past utterances. This subset captures salient contextual-aware user traits for downstream simulation.

\paragraph{Pseudo Utterance Generation (Roughly)}  
To simulate user-specific expression patterns, we employ a large language model $\mathcal{G}^{\text{utt}}_{\text{LLM}}$ to generate $m$ pseudo utterances conditioned on the original query $q$ and user dialogue history $H_q$. Each utterance $f_i$ is designed to reflect how the user might naturally articulate the query, capturing stylistic nuances such as tone, verbosity, and phrasing:
\begin{equation}
F = \{f_1, \dots, f_m\} = \mathcal{G}^{\text{utt}}_{\text{LLM}}(q, H_q), \quad \mathbf{f}_{\text{avg}} = \frac{1}{m} \sum_{i=1}^{m} \phi(f_i),
\end{equation}
where $\phi(f_i)$ denotes the embedding of the $i$-th utterance and $\mathbf{f}_{\text{avg}}$ is their mean representation.

\paragraph{Pseudo Reasoning Generation (Logically)}  
Beyond surface expression, effective personalization also requires modeling the user’s implicit reasoning process. We define $\mathcal{G}^{\text{rea}}_{\text{LLM}}$ as a parallel simulating pipeline to elicit a step-by-step rationale $r$ of user expression:
\begin{equation}
r = \mathcal{G}^{\text{rea}}_{\text{LLM}}(q, H_q), \quad \mathbf{r} = \phi(r).
\end{equation}
This rationale introduces logical cues that are often absent from surface queries, thereby providing a complementary semantic signal to guide expansion alignment.

\begin{table*}[t]
\centering
\small
\setlength\tabcolsep{2.75pt}  
\renewcommand{\arraystretch}{0.80} 
\begin{tabular}{lcccccccccccccc}
\toprule
 \textbf{Method} & \multicolumn{2}{c}{\textbf{Base}} & \multicolumn{2}{c}{\textbf{HyDE}} & \multicolumn{2}{c}{\textbf{Query2Term}} & \multicolumn{2}{c}{\textbf{MILL}} & \multicolumn{2}{c}{\textbf{CoT}} & \multicolumn{2}{c}{\textbf{ThinkQE}} & \multicolumn{2}{c}{\textbf{PBR (Ours)}} \\
\cmidrule(r){2-3} \cmidrule(r){4-5} \cmidrule(r){6-7} \cmidrule(r){8-9} \cmidrule(r){10-11} \cmidrule(r){12-13} \cmidrule(r){14-15}
Metrics & R@5 & N@5 & R@5 & N@5 & R@5 & N@5 & R@5 & N@5 & R@5 & N@5 & R@5 & N@5 & R@5 & N@5 \\
\midrule
\rowcolor{gray!15}
\multicolumn{15}{c}{\textit{multi-qa-MiniLM-L6-cos-v1}} \\
\textbf{Overall} & 0.4484 & 0.3669 & 0.3464 & 0.2945 & 0.3584 & 0.3060 & 0.3200 & 0.3254 & 0.3627 & 0.3000 &\underline{0.4791}&\underline{0.3902}& \textbf{0.5035} & \textbf{0.4201} \\
Basic information & 0.4515 & 0.3088 & 0.3106 & 0.2395 & 0.3424 & 0.2614 & 0.2879 & 0.2337 & 0.3424 & 0.2454 &\underline{0.4606}&\underline{0.3265}& \textbf{0.5091} & \textbf{0.3647} \\
Preference (hard) & 0.3659 & 0.3759 & 0.3122 & 0.3079 & 0.3659 & 0.3491 & 0.2927 & \underline{0.4250} & 0.3171 & 0.3335 &\textbf{0.4195}&\textbf{0.4310}& \underline{0.4049} & 0.4175 \\
Social & 0.4852 & 0.4356 & 0.3909 & 0.3259 & 0.3494 & 0.3144 & 0.3808 & 0.3923 & 0.4009 & 0.3320 &\underline{0.5503}&\underline{0.4554}& \textbf{0.5541} & \textbf{0.4914} \\
Preference (easy) & 0.4904 & 0.4582 & 0.4615 & 0.4422 & 0.4327 & 0.4095 & 0.3750 & 0.4195 & 0.4423 & 0.4133 &\underline{0.5064}&\underline{0.4626}& \textbf{0.5321} & \textbf{0.5129} \\
\midrule
\rowcolor{gray!15}
\multicolumn{15}{c}{\textit{all-MiniLM-L6-v2}} \\
\textbf{Overall} & 0.3783 & 0.3074 & 0.3747 & 0.3144 & 0.3908 & 0.3256 & 0.3030 & 0.3059 & 0.3721 & 0.3098 &\underline{0.3861}&\underline{0.339}3& \textbf{0.4516} & \textbf{0.3855} \\
Basic information & 0.3515 & 0.2644 & \underline{0.4015} & \underline{0.2975} & 0.3894 & 0.2914 & 0.3030 & 0.2547 & 0.3455 & 0.2696 &0.3455&0.2677& \textbf{0.4515} & \textbf{0.3485} \\
Preference (hard) & \underline{0.4000} & 0.3547 & 0.3512 & 0.3561 & 0.3805 & 0.3798 & 0.3220 & \underline{0.4171} & 0.3463 & 0.3490 &0.3366&0.3486& \textbf{0.4341} & \textbf{0.4352} \\
Social & 0.3921 & 0.3048 & 0.2805 & 0.2435 & 0.3984 & 0.3163 & 0.2638 & 0.2829 & 0.4160 & 0.3046 &\textbf{0.4780}&\textbf{0.4334}& \underline{0.4494} & \underline{0.3777} \\
Preference (easy) & 0.4295 & 0.4199 & \textbf{0.4904} & \underline{0.4646} & 0.3974 & 0.4038 & 0.3526 & 0.3944 & 0.4359 & 0.4285 &0.4487&0.4357& \underline{0.4840} & \textbf{0.4800} \\
\midrule
\rowcolor{gray!15}
\multicolumn{15}{c}{\textit{bge-base-en-v1.5}} \\
\textbf{Overall} & \underline{0.3738} & 0.3015 & 0.3108 & 0.2597 & 0.3007 & 0.2497 & 0.2791 & 0.2869 & 0.3199 & 0.2642 &0.3643&\underline{0.3156}& \textbf{0.4029} & \textbf{0.3402} \\
Basic information & \underline{0.3970} & 0.2748 & 0.2955 & 0.2051 & 0.3000 & 0.1955 & \underline{0.2879} & 0.2319 & 0.3152 & 0.2052 &0.3152&0.2430& \textbf{0.4121} & \textbf{0.3057} \\
Preference (hard) & 0.3268 & 0.3343 & 0.3073 & 0.3148 & 0.2976 & 0.3186 & 0.2244 & 0.3566 & 0.3073 & 0.3281 &\underline{0.3463}&\underline{0.3635}& \textbf{0.3707} & \textbf{0.3889} \\
Social & 0.3204 & 0.2799 & 0.2714 & 0.2443 & 0.2852 & 0.2503 & 0.2752 & 0.3004 & 0.2965 & 0.2647 &\textbf{0.4261}&\textbf{0.3735}& \underline{0.3657} & \underline{0.3089} \\
Preference (easy) & 0.4583 & 0.4065 & 0.4615 & \underline{0.4356} & 0.3397 & 0.3693 & 0.3365 & 0.3819 & 0.4071 & 0.4117 &\underline{0.4744}&0.4289& \textbf{0.4904} & \textbf{0.4734} \\
\midrule
\textbf{Overall Average} & 0.4002 & 0.3253 & 0.3440 & 0.2895 & 0.3500 & 0.2938 & 0.3007 & 0.3061 & 0.3295 & 0.3121 &\underline{0.4098}& \underline{0.3484}& $\textbf{0.4527}^*$ & $\textbf{0.3819}^*$\\
\bottomrule
\end{tabular}
\caption{Retrieval performance in \textbf{PersonaBench}. The best results are in \textbf{bold}, and the second-best results are \underline{underlined}. ``\textbf{{\Large *}}'' indicates the statistically significant improvements (i.e., two-sided t-test with $p<0.05$) over the best baseline. }
\label{tab:main_results}
\vspace{-10pt}
\end{table*}

\subsection{P-Anchor}
Existing methods are often designed to retrieve from a unified RAG corpus, lacking explicit grounding in the structural organization of individual user corpora~\cite{tan2025personabench,wu2025longmemeval}.
To address this, we propose \textbf{P-Anchor}, a structure-aware module that captures corpus-level preferences by identifying semantically central regions within a graph-structured user space.

\paragraph{Graph Construction from User Corpus.}
We represent the user's prior corpus—such as interaction records with AI assistants—as a semantic graph $G = (\mathcal{C}, \mathcal{E})$, following previous study~\cite{tang2025llm4tag}, where $\mathcal{C}$  and $\mathcal{E}$ denote the sets of nodes and edges, respectively. 
The set of nodes $\mathcal{C}$ consists of history corpus vectors, where each node $\mathbf{c}_i$ as in Eq. \eqref{eq:c_i} represents an encoded segment. The edges $\mathcal{E}$ are established based on pairwise similarity between the nodes.
For each pair, we compute cosine similarity:
\begin{equation}
S_{ij} = \frac{\langle \mathbf{c}_i, \mathbf{c}_j \rangle}{\|\mathbf{c}_i\| \cdot \|\mathbf{c}_j\|}.
\end{equation}
We construct a sparse adjacency matrix $A$ to represent the structural links within the user corpus, which retains only the top-$k_2$ neighbors exceeding a similarity threshold $\theta$:
\begin{equation}
A_{ij} = 
\begin{cases}
S_{ij}, & \text{if } S_{ij} \geq \theta \text{ and } j \in \text{top-}k_2(i), \\
0, & \text{otherwise}.
\end{cases}
\end{equation}

\paragraph{Graph-Based Anchor Representation.}
We following PageRank~\cite{page1999pagerank,haveliwala2002topic}  over $G$ to estimate a stationary distribution $\boldsymbol{\pi}$ reflecting node centrality. Let $\mathbf{P}=norm(\mathbf{A})$ be the row-normalized transition matrix, then:
\begin{equation}
\boldsymbol{\pi}^{(t+1)} = \alpha \cdot \mathbf{P}^\top \boldsymbol{\pi}^{(t)} + (1 - \alpha) \cdot \frac{1}{n} \mathbf{1}.
\end{equation}
The user anchor is computed as a weighted aggregation:
\begin{equation}
\mathbf{c}_{\text{Anchor}} = \sum_{i=1}^{n} \boldsymbol{\pi}_i \cdot \mathbf{c}_i.
\end{equation}
This representation encodes structurally central semantics from the user's corpus, anchoring the query in a context-aware and personalized semantic space.

\subsection{PBR Fusion}  

To synthesize the signals captured in P-PRF and P-Anchor, we design a fusion mechanism that integrates stylistic pseudo-feedback with the query and user anchor. Recognizing that the utility of pseudo-feedback varies across users and queries—depending on individual expression patterns and the semantic openness of the query—we introduce a \textit{personalized and query-specific dynamic weight} that balances the contribution of P-PRF according to its semantic proximity to both the original query $\mathbf{q}$ and the corpus anchor $\mathbf{c}_{\text{Anchor}}$. Specifically, we compute two weights $w_1$ and $w_2$ to quantify the semantic alignment of the pseudo-utterance and reasoning feedback, respectively:
\begin{equation}
w_1 = 1 + \text{sim}\left( \frac{\mathbf{q} + \mathbf{c}_{\text{Anchor}}}{2}, \mathbf{f}_{\text{avg}} \right), \quad
\end{equation}
\begin{equation}
w_2 = 1 + \text{sim}\left( \frac{\mathbf{q} + \mathbf{c}_{\text{Anchor}}}{2}, \mathbf{r} \right),
\end{equation}
where $\text{sim}(\cdot)$ denotes cosine similarity. The additive constant ensures positivity and smooth interpolation.

Then, we define a user-specific personalization shift as:
\begin{equation}
\Delta_\text{user}(\mathbf{q}, \mathcal{C}) = \mathbf{c}_{\text{Anchor}} + w_1 \cdot \mathbf{f}_{\text{avg}} + w_2 \cdot \mathbf{r},
\end{equation}
and construct the final personalized query embedding as:
\begin{equation}
\mathbf{q}^* = \mathbf{q} + \Delta_\text{user}(\mathbf{q}, \mathcal{C}).
\end{equation}

The resulting query $\mathbf{q}^*$ incorporates user-specific semantics from both style and structure. It encodes not only the immediate surface form of the input query, but also latent stylistic preferences, goal-driven reasoning, and historically salient corpus signals.

\paragraph{Final Retrieval.}  
We retrieve from the user corpus $\mathcal{C}$ using \texttt{faiss}-based~\cite{douze2024faiss} nearest neighbor search with the fused query vector $\mathbf{q}^*$. This process leverages embedding-level alignment to recover user-relevant content.

\begin{table*}[ht]
\centering
\small
\setlength\tabcolsep{2.2pt}  
\renewcommand{\arraystretch}{0.80} 
\begin{tabular}{lcccccccccccc}
\toprule
 \textbf{Dataset} & \multicolumn{6}{c}{\textbf{LongMenEval-s}} & \multicolumn{6}{c}{\textbf{LongMenEval-m}} \\
 \cmidrule(lr){2-7} \cmidrule(lr){8-13}
Metrics & R@1 & N@1 & R@3 & N@3 & R@5 & N@5 & R@1 & N@1 & R@3 & N@3 & R@5 & N@5 \\
\midrule
\rowcolor{gray!15}
\multicolumn{13}{c}{\textit{multi-qa-MiniLM-L6-cos-v1}} \\
Base     & 0.1885 & 0.7924 & 0.6730 & 0.8020 & 0.8067 & 0.8301 & 0.1098 & \underline{0.5251} & 0.4057 & 0.5559 & 0.5394 & 0.6038 \\
HyDE     & \underline{0.1957} & 0.7852 & 0.6945 & 0.8173 & 0.8115 & \underline{0.8436} & 0.1146 & 0.4964 & 0.4129 & 0.5411 & 0.5179 & 0.5828 \\
Query2Term & 0.1814 & 0.7446 & 0.6396 & 0.7684 & 0.7470 & 0.7959 & 0.1146 & 0.4749 & 0.3747 & 0.5091 & 0.4821 & 0.5531 \\
MILL     & 0.1838 & 0.7709 & 0.6635 & 0.7845 & 0.6635 & 0.7712 & 0.1122 & 0.5155 & 0.4129 & 0.5552 & 0.4129 & 0.5468 \\
CoT      & 0.1862 & 0.7470 & 0.6516 & 0.7707 & 0.7542 & 0.8006 & 0.1098 & 0.4654 & 0.3866 & 0.5110 & 0.4726 & 0.5475 \\
ThinkQE & 0.1933 & \underline{0.8019} & \underline{0.7064} & \underline{0.8214} & \underline{0.8162 }& 0.8424 & \underline{0.1169} & 0.5227 & \underline{0.4168} & \underline{0.5647} & \underline{0.5517} & \underline{0.6156} \\
\textbf{PBR} & $\mathbf{0.2100^*}$ & $\mathbf{0.8210^*}$ & $\mathbf{0.7279^*}$ & $\mathbf{0.8444^*}$ & $\mathbf{0.8282^*}$ & $\mathbf{0.8598^*}$ & $\mathbf{0.1217^*}$ & $\mathbf{0.5537^*}$ & $\mathbf{0.4320^*}$ & $\mathbf{0.5776^*}$ & $\mathbf{0.5537^*}$ & $\mathbf{0.6177^*}$ \\
\midrule
\rowcolor{gray!15}
\multicolumn{13}{c}{\textit{all-MiniLM-L6-v2}}\\
Base     & \underline{0.2196} & \underline{0.8234} & \underline{0.7399} & \underline{0.8477} & \underline{0.8568} & \underline{0.8731} & \underline{0.1265} & 0.5322 & 0.4415 & 0.5857 & 0.5346 & 0.6191 \\
HyDE     & 0.2172 & 0.8138 & 0.7208 & 0.8372 & 0.8377 & 0.8599 & 0.1074 & 0.4988 & 0.4081 & 0.5483 & 0.5131 & 0.5884 \\
Query2Term & \underline{0.2196} & 0.8019 & 0.6778 & 0.8000 & 0.8138 & 0.8351 & 0.1193 & 0.4940 & 0.3747 & 0.5209 & 0.4964 & 0.5689 \\
MILL     & 0.2100 & 0.8162 & 0.7208 & 0.8382 & 0.7208 & 0.8257 & 0.1217 & 0.5179 & 0.3962 & 0.5530 & 0.3962 & 0.5450 \\
CoT      & 0.2124 & 0.7709 & 0.6730 & 0.7932 & 0.7780 & 0.8187 & 0.1193 & 0.4487 & 0.3389 & 0.4920 & 0.4344 & 0.5273 \\
ThinkQE & \underline{0.2196} & \underline{0.8234} & 0.7232 & 0.8332 & 0.8162 & 0.8527 & 0.1241 & \underline{0.5585} & \underline{0.4461} & \underline{0.5871} & \underline{0.5537} & \underline{0.6229} \\
\textbf{PBR} & $\mathbf{0.2267^*}$ & $\mathbf{0.8592^*}$ & $\mathbf{0.7780^*}$ & $\mathbf{0.8754^*}$ & $\mathbf{0.8640^*}$ & $\mathbf{0.8902^*}$ & $\mathbf{0.1408^*}$ & $\mathbf{0.5800^*}$ & $\mathbf{0.4463^*}$ & $\mathbf{0.5949^*}$ & $\mathbf{0.5847^*}$ & $\mathbf{0.6424^*}$ \\
\midrule
\rowcolor{gray!15}
\multicolumn{13}{c}{\textit{bge-base-en-v1.5}} \\
Base     & \underline{0.2267} & 0.8616 & 0.7900 & 0.8863 & 0.8926 & 0.9007 & 0.1217 & 0.5943 & \underline{0.5322} & \underline{0.6605} & 0.6444 & 0.6929 \\
HyDE     & 0.2196 & 0.8759 & 0.7995 & 0.8974 & 0.8998 & 0.9044 & 0.1313 & 0.5943 & 0.5298 & 0.6538 & \underline{0.6659} & 0.7010 \\
Query2Term & 0.2100 & 0.8544 & 0.7637 & 0.8739 & 0.8926 & 0.8989 & 0.1360 & 0.5776 & 0.5155 & 0.6290 & 0.6348 & 0.6754 \\
MILL     & \underline{0.2267} & 0.8663 & 0.7804 & 0.8819 & 0.7804 & 0.8680 & 0.1289 & 0.5967 & 0.5107 & 0.6362 & 0.5107 & 0.6278 \\
CoT      & 0.2196 & 0.8425 & 0.7637 & 0.8707 & 0.8783 & 0.8918 & \underline{0.1384} & 0.5609 & 0.4964 & 0.6148 & 0.5967 & 0.6570 \\
ThinkQE & 0.2124 & \underline{0.8807} & \underline{0.8108} & \underline{0.9012} & \underline{0.9021} & \underline{0.9126} & 0.1384 & \underline{0.6072} & 0.5080 & 0.6321 & 0.6445 & \underline{0.7013} \\
\textbf{PBR} & $\mathbf{0.2315^*}$ & $\mathbf{0.8902^*}$ & $\mathbf{0.8138^*}$ & $\mathbf{0.9072^*}$ & $\mathbf{0.9045^*}$ & $\mathbf{0.9191^*}$ & $\mathbf{0.1408^*}$ & $\mathbf{0.6205^*}$ & $\mathbf{0.5489^*}$ & $\mathbf{0.6612^*}$ & $\mathbf{0.6874^*}$ & $\mathbf{0.7060^*}$\\
\bottomrule
\end{tabular}
\caption{Retrieval performance in \textbf{LongMemEval}. The best results are in \textbf{bold}, and the second-best results are \underline{underlined}. ``\textbf{{\Large *}}'' indicates the statistically significant improvements (i.e., two-sided t-test with $p<0.05$) over the best baseline. }
\label{tab:longmen_results}
\vspace{-10pt}
\end{table*}

\section{Experiment}

In this section, we conduct experiments to address the following research questions:
\begin{itemize}[leftmargin=*]
    \item \textbf{RQ1}: \emph{Does the proposed PBR method improve retrieval performance on personalized corpus?}
    \item \textbf{RQ2}: \emph{What are the individual contributions of the P-PRF and P-Anchor modules to the overall performance?}
    \item \textbf{RQ3}: \emph{How does information propagation within the P-Anchor module affect the alignment between the query and user-specific corpora?}
\end{itemize}

\subsection{Experiment Setting}

\paragraph{Datasets.}
We evaluate our method on two recent benchmarks tailored for personalized retrieval. \textbf{PersonaBench}~\cite{tan2025personabench} simulates user-specific queries over synthetic private data (e.g., preferences, behaviors, demographics), requiring models to align responses with personalized context. It includes both explicit and implicit user traits across diverse domains. \textbf{LongMemEval}~\cite{wu2025longmemeval} targets long-term interactive memory, where each user is associated with multi-turn dialogue traces. We use two subsets: \textit{LongMemEval-s} (sparse history) and \textit{LongMemEval-m} (dense history), to evaluate the effectiveness of retrieval under both short- and long-horizon personalization. More details are provided in Appendix 1.

\paragraph{Evaluation Metrics.}
We report performance using two standard retrieval metrics: \textbf{Recall@K (R@K)}, which measures the proportion of relevant items ranked within the top-K retrieved results, and \textbf{NDCG@K (N@K)} to reflect retrieval quality.

\paragraph{Baselines.}
We compare our method with six strong baselines: \textbf{Base}, a static query retriever without expansion; \textbf{HyDE}~\cite{gao2023precise}, which generates hypothetical answers to guide retrieval; \textbf{Query2Term}~\cite{jagerman2023query}, which performs keyphrase-based query expansion; \textbf{MILL}~\cite{jia2023mill}, which leverages multi-level logical structures; \textbf{CoT}~\cite{wei2022chain}, which applies chain-of-thought reasoning for better ranking; and \textbf{ThinkQE}~\cite{lei2025thinkqe}. 

\paragraph{Implement details.}
To verify the robustness of PBR, all methods are evaluated under three retrieval backbones: \texttt{multi-qa-MiniLM-L6-cos-v1}~\cite{wang2020minilm}, \texttt{all-MiniLM-L6-v2}~\cite{wang2020minilm}, and \texttt{bge-base-en-v1.5}~\cite{chen2024bge}. For a fair comparison, all methods use an advanced LLM and run 5 times. In \textbf{PBR} framework, we set $k_1=5$,$m=5$, $\theta=0.75$, and $k_2=10$ across \textit{Perosnabench} and \textit{LongMemEval-s}. The larger $k_2=50$ is particularly suited for \textit{LongMemEval-m}, which contains richer user histories. More details are provided in Appendix 2.

\subsection{Overall Performance (RQ1)}

\begin{figure*}[ht]
    \centering
    \includegraphics[width=0.9\linewidth]{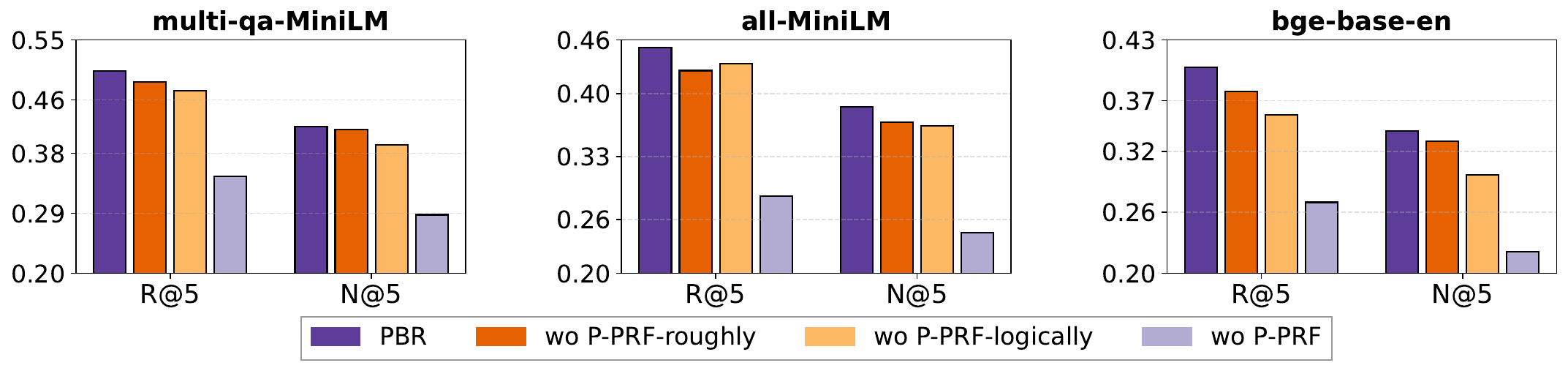}
    \caption{%
        Overall performance in \textbf{PersonaBench} comparison across three retrieval models. 
    }
    \label{fig:pprf-ablation}
    \vspace{-10pt}
\end{figure*}

\begin{figure*}[ht]
    \centering
    \includegraphics[width=1.0\linewidth]{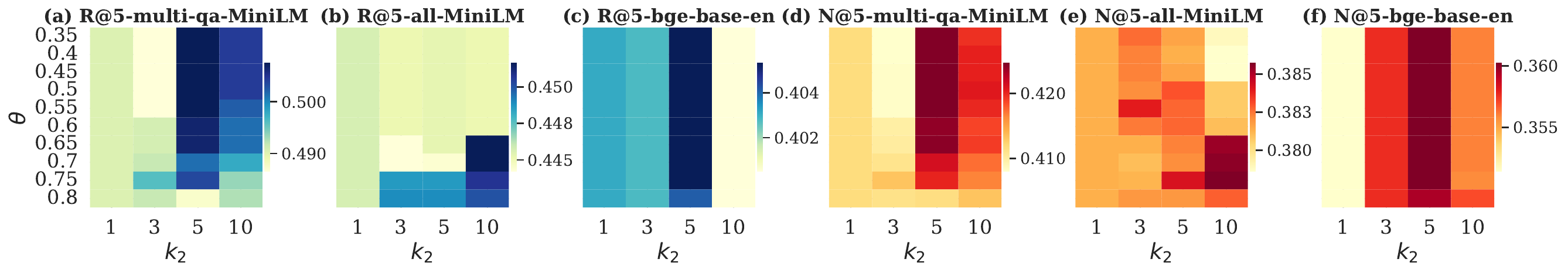}
    \caption{Parameter sensitivity analysis on P-Anchor of $\theta$ and $k_2$ settings.}
    \label{fig:anchor_sensitivity}
    \vspace{-10pt}
\end{figure*}

To comprehensively evaluate the effectiveness of PBR, we conduct experiments on two benchmarks. \textbf{PersonaBench} contains 6 users, each with a dedicated corpus and about 50 user queries characterized by ambiguous intent and highly personalized language. In contrast, \textbf{LongMemEval} simulates a large-scale factual QA scenario with 500 distinct user queries, each paired with a relevant memory corpus. The former highlights soft personalization challenges (e.g., preferences and social traits), while the latter emphasizes hard factual grounding under long-context retrieval.

\textbf{PBR resolving semantic ambiguity through user-specific expansion.} 
Queries in PersonaBench often exhibit semantic openness and personal references—such as ``Where is my hometown?'' or ``What is my favorite color?''—which cannot be resolved without grounding in the user's unique corpus. As shown in Table~\ref{tab:main_results}, our PBR framework achieves the best overall performance across all retrievers (R@5: \textbf{0.4527}, N@5: \textbf{0.3819}), significantly surpassing the best baseline (ThinkQE: \underline{0.4098} / \underline{0.3484}). This gain stems from PBR’s ability to generate user-style pseudo-feedback that anchors the query into a meaningful semantic space, reducing ambiguity and improving retrieval accuracy. 

\textbf{PBR excels in factual retrieval by boosting top-ranked precision.} 
In contrast to PersonaBench, queries in LongMemEval are fact-oriented with specific personalization requirements, each associated with a specific memory corpus. In this setting, PBR boosts top-ranked retrieval performance. In Table~\ref{tab:longmen_results}, PBR also outperforms all baselines on LongMemEval, particularly on R@1, where it achieves \textbf{0.2315} on the -s version and \textbf{0.1408} on the -m version using \texttt{bge}. Compared to the best-performing baseline (0.2267 / 0.1384), PBR achieves absolute gain on top-1 precision. This suggests that our style simulation and structural anchoring more precisely locate relevant memory, elevating correct answers to top-1 even in long-context settings.

\subsection{Ablation Study (RQ2)}
To investigate the individual roles of \textbf{P-PRF} and \textbf{P-Anchor} in our framework, we perform a module-wise ablation across three retrievers in PersonaBench, as shown in Table~\ref{tab:ablation_results}. Each model variant is evaluated under both R@5 and N@5. More details are provided in Appendix 3.

\begin{table}[ht]
\centering
\renewcommand{\arraystretch}{0.90} 
\resizebox{85mm}{!}{
\begin{tabular}{lcccccc}
\toprule
& \multicolumn{2}{c}{\textbf{PBR}} 
& \multicolumn{2}{c}{\textbf{w/o P-Anchor}} 
& \multicolumn{2}{c}{\textbf{w/o P-PRF}}\\
\cmidrule(lr){2-3} \cmidrule(lr){4-5} \cmidrule(lr){6-7} 
Metrics & R@5 & N@5 & R@5 & N@5 & R@5 & N@5   \\
\midrule
\rowcolor{gray!15}
\multicolumn{7}{c}{\textit{multi-qa-MiniLM-L6-cos-v1}} \\
Overall & \textbf{0.5035} & \textbf{0.4201} & 0.4871 & 0.4100 & 0.3457 & 0.2877 \\
Basic information & \textbf{0.5091} & \textbf{0.3647} & 0.4833 & 0.3570 & 0.3318 & 0.2242 \\
Preference (hard) & 0.4049 & 0.4175 & \textbf{0.4390} & \textbf{0.4295} & 0.2683 & 0.2981  \\
Social & \textbf{0.5541} & \textbf{0.4914} & 0.5164 & 0.4529 & 0.4277 & 0.3851 \\
Preference (easy) & \textbf{0.5321} & \textbf{0.5129} & 0.5192 & 0.5160 & 0.3590 & 0.3414 \\
\midrule
\rowcolor{gray!15}
\multicolumn{7}{c}{\textit{all-MiniLM-L6-v2}}\\
Overall & \textbf{0.4516} & \textbf{0.3855} & 0.4382 & 0.3729 & 0.2860 & 0.2449 \\
Basic information & \textbf{0.4515} & \textbf{0.3485} & 0.4379 & 0.3393 & 0.1985 & 0.1617 \\
Preference (hard) & \textbf{0.4341} & \textbf{0.4352} & 0.4244 & 0.4201 & 0.2878 & 0.2718 \\
Social & \textbf{0.4494} & \textbf{0.3777} & 0.4333 & 0.3639 & 0.4179 & 0.3387 \\
Preference (easy) & \textbf{0.4840} & \textbf{0.4800} & 0.4712 & 0.4587 & 0.3846 & 0.3634  \\
\midrule
\rowcolor{gray!15}
\multicolumn{7}{c}{\textit{BAAI/bge-base-en-v1.5}} \\ 
Overall & \textbf{0.4029} & \textbf{0.3402} & 0.3921 & 0.3256 & 0.2699 & 0.2214  \\
Basic information & \textbf{0.4121} & \textbf{0.3057} & 0.3970 & 0.2958 & 0.2576 & 0.1804 \\
Preference (hard) & 0.3707 & \textbf{0.3889} & \textbf{0.3902} & 0.3778 & 0.2146 & 0.2173  \\
Social & \textbf{0.3657} & \textbf{0.3089} & 0.3431 & 0.2822 & 0.3261 & 0.2816  \\
Preference (easy) & \textbf{0.4904} & \textbf{0.4734} & 0.4744 & 0.4582 & 0.2949 & 0.2785  \\
\midrule
\textbf{Average} & \textbf{0.4527} & \textbf{0.3819} & 0.4391 & 0.3695 & 0.3005 & 0.2513 \\
\bottomrule
\end{tabular}
}
\caption{Ablation study of PBR in \textbf{PersonaBench}.}
\label{tab:ablation_results}
\vspace{-10pt}
\end{table}

\textbf{P-PRF drives major performance gains in personalized retrieval.}
Removing P-PRF leads to a substantial performance drop across all retrievers (e.g., on all-MiniLM, R@5 falls from 0.4516 to 0.2860), with the most severe decline observed on personalization-heavy subsets. This underscores the importance of P-PRF in generating semantically rich and stylistically aligned pseudo queries that effectively capture user intent.

\textbf{Effectiveness of P-PRF Components.}
To assess how \textbf{P-PRF} enhances intent modeling, we perform ablation studies in PersonaBench and three retrievers (Figure~\ref{fig:pprf-ablation}). Removing either the \textit{roughly}-aligned pseudo-utterance or \textit{logically}-structured pseudo-reasoning consistently degrades performance in R@5 and N@5. The former harms lexical coverage, while the latter impairs goal-oriented reasoning, reducing ranking precision on inference-heavy datasets.

\textbf{P-Anchor promotes semantic alignment in structured corpora.}
Removing P-Anchor leads to a consistent drop (e.g., N@5 from \textbf{0.3819} to 0.3695), especially in \textit{Basic Information} and \textit{Social}, where user corpora exhibit clearer structure. In contrast, in non-clustered tasks like \textit{Preference (hard)}, it offers limited gain and may overfit. P-Anchor is most effective when user context is semantically coherent.

\subsection{Parameter Sensitivity Analysis (RQ3)}

\begin{figure*}[ht]
  \centering
  \begin{subfigure}[t]{0.32\textwidth}
    \includegraphics[width=1.0\linewidth]{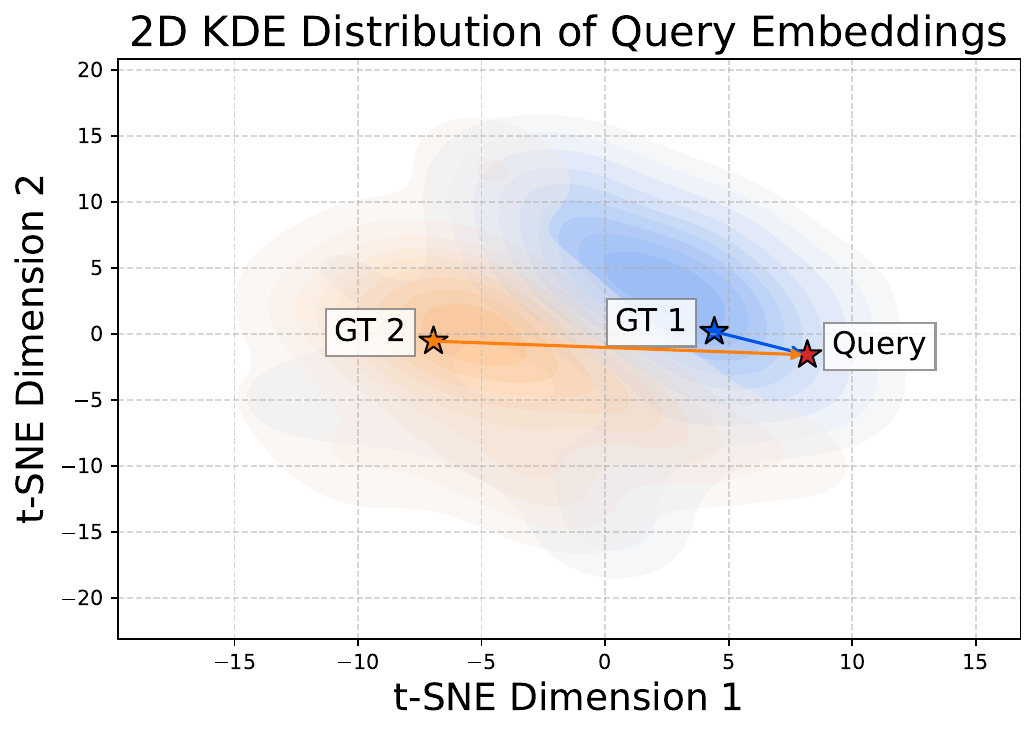}
    \caption{Query-GT Comparison}
    \label{fig:query_gt}
  \end{subfigure}
  \hfill
  \begin{subfigure}[t]{0.32\textwidth}
    \includegraphics[width=1.0\linewidth]{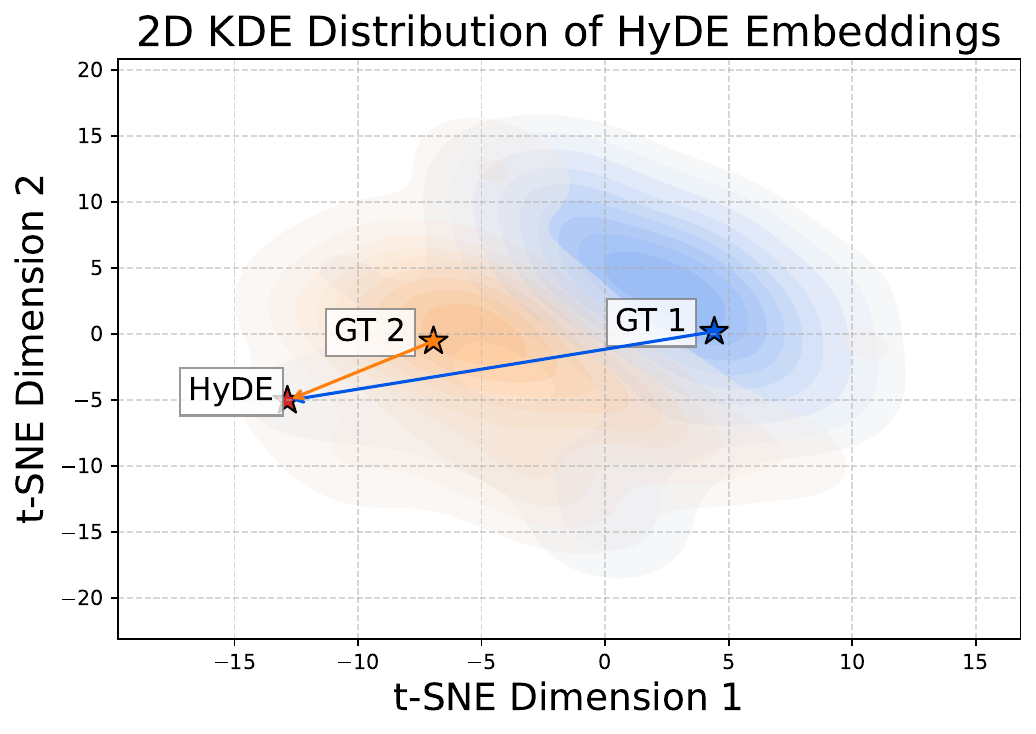}
    \caption{HyDE-GT Comparison}
    \label{fig:hyde_gt}
  \end{subfigure}
  \hfill
  \begin{subfigure}[t]{0.32\textwidth}
    \includegraphics[width=1.0\linewidth]{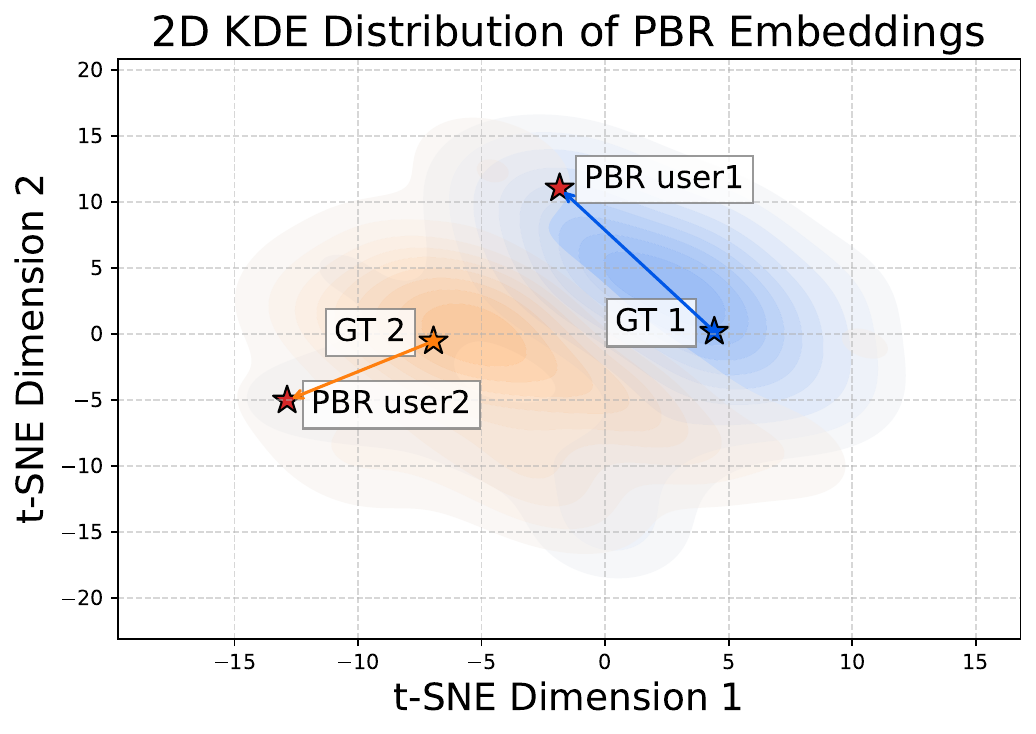}
    \caption{PBR-GT Comparison (Ours)}
    \label{fig:ours_gt}
  \end{subfigure}
  \caption{Visualization of retrieval distance of the query in the Figure~\ref{fig:intro} compared to ground truth (GT) under different methods: (a) vanilla query, (b) HyDE, and (c) our proposed PBR.}
  \label{fig:retrieval_gt_comparison}
  \vspace{-10pt}
\end{figure*}

To verify how information propagation within the P-Anchor module influences the alignment between queries and user-specific semantic space, we conduct a sensitivity analysis over two structural hyperparameters: the semantic threshold $\theta$ and the neighbor size $k_2$. The results are shown in Figure~\ref{fig:anchor_sensitivity}.

\textbf{Moderate propagation improves alignment.} When $\theta \in [0.65, 0.75]$ and $k_2 = 5$, both R@5 and N@5 reach consistent peaks across datasets, indicating that controlled expansion of semantic neighborhoods facilitates better anchoring and personalization.

\textbf{Over-propagation causes semantic drift.} As $\theta$ and $k_2$ continue to increase, performance either saturates or slightly declines (e.g., at $\theta=0.8$, $k_2=10$), suggesting that excessive propagation introduces noise and weakens the semantic specificity of the alignment.

\subsection{Visualization Case Study}

To verify the practical effectiveness of our proposed PBR framework, we conduct a retrieval visualization case study.
Figure~\ref{fig:retrieval_gt_comparison} visualizes the t-SNE distribution of user queries, ground truths (GT), and generated queries under different methods.
The results show that traditional expansion methods produce generic results misaligned with different users.
Measuring cosine similarity to user-specific ground truth, PBR achieves higher separation between users (0.31 vs. 0.27), outperforming HyDE (0.04 vs. –0.01) and the original query (0.05 vs. 0.10). This highlights PBR’s effectiveness in embedding user-personalized semantics.

\textbf{P-PRF enables semantically diverse and intent-aligned queries.}
Compared to vanilla queries or HyDE-generated expansions, PBR-generated queries are semantically closer to ground truth responses in both user clusters.
The trajectory of PBR vectors more accurately captures user-relevant preferences, confirming its effectiveness in generating expressive and personalized query variants.

\textbf{Improved alignment leads to more successful retrieval.}
Unlike HyDE, which generates a generic QE of all users, PBR yields multiple expansions (e.g., PBR user1 and PBR user2) that better cover the user-specific semantic space.
This personalized QE allows the retriever to more precisely locate relevant results near each GT, reducing retrieval distance and improving semantic alignment.

\FloatBarrier
\section{Literature Review}

\subsection{Query Expansion in RAG System}

Query expansion (QE) improves retrieval performance by simulating PRF. With the rise of LLMs, recent studies have explored LLMs as zero-shot QE engines. HyDE~\cite{gao2023precise} and Query2Doc~\cite{wang2023query2doc} demonstrate that prompting LLMs can yield precise expansions with PRF. Other works enhance semantic diversity or reasoning depth: MILL~\cite{jia2023mill} employ mutual verification, and GRF~\cite{mackie2023generative} apply LLM-based relevance feedback. ThinkQE~\cite{lei2025thinkqe} models expansion as a reasoning process, while LLMlingua~\cite{jiang2023llmlingua} focuses on compressing expansion prompts. GAR~\cite{xia2024knowledge} further introduces a knowledge graph-aware QE framework.

However, existing work overlooks personalization and fails to align QE with user-specific semantic spaces. In contrast, our PBR generates style-aware and structure-aware expansions, bridging this gap for personalized retrieval.

\subsection{Personalized RAG System}

Personalized RAG system has recently emerged as a crucial direction for adapting LLMs to individual users. A recent survey by Li et al.~\cite{li2025survey,li2023agent4ranking} provides a comprehensive overview, highlighting the importance of personalization in RAG. Approaches such as UniMS-RAG~\cite{wang2024unims} integrate multi-source signals for dialogue personalization, while PersonaRAG~\cite{zerhoudi2024personarag} enhances personalization via role-based agent frameworks. Salemi et al.~\cite{salemi2024lamp} further investigate LLM personalization by modeling language preferences and user goals. Complementary work like Wu et al.~\cite{wu2024understanding} explores the role of explicit user profiles, and Cohn et al.~\cite{cohn2025personalizing} leverage interaction logs for educational agent adaptation. Domain-specific applications, such as personalized care systems~\cite{yang2025improving} and memory-based assistant evaluation~\cite{wu2025longmemeval,xu2025towards,xu2025align}, highlight the importance of long-term memory and evolving user states in effective RAG.

Despite recent advances, little attention has been given to improving retrieval accuracy at the pre-retrieval stage. PBR enables fine-grained personalization by jointly modeling the user’s style and corpus structure before retrieval.

\section{Conclusion}

This paper introduces \textbf{PBR}, a \textit{Personalize-Before-Retrieve} framework for personalized retrieval. PBR expands user query prior to retrieval by integrating two components: \textbf{P-PRF}, a style-aware pseudo query generator, and \textbf{P-Anchor}, a graph-based corpus alignment module. 
Our results highlight the importance of pre-retrieval personalization for improving user-centric retrieval performance.
Future work will investigate generalizing PBR to broader information-seeking tasks beyond query-based retrieval.

\section*{Acknowledgments}
This research was supported by the National Natural Science Foundation of China (NSFC) under Grant 72071029, 72231010, 62502404, and the Graduate Research Fund of the School of Economics and Management of Dalian University of Technology (No. DUTSEMDRFKO1). This research was partially supported by Hong Kong Research Grants Council (Research Impact Fund No.R1015-23, Collaborative Research Fund No.C1043-24GF, General Research Fund No.11218325), Institute of Digital Medicine of City University of Hong Kong (No.9229503), and Huawei (Huawei Innovation Research Program).

\nocite{*}

\bibliography{chapter_new/ref_final}

\section*{Appendix 1: Datasets Details}

\paragraph{PersonaBench.} 
PersonaBench~\cite{tan2025personabench}\footnote{\url{https://github.com/SalesforceAIResearch/personabench}} is a benchmark dataset designed to evaluate personalized retrieval grounded in user-specific contexts. It includes six users, each associated with a heterogeneous personal corpus composed of (i) conversations with friends (Conv.), (ii) dialogues with AI assistants (AI.), and (iii) e-commerce purchase histories (E-com.). For each user, 40--50 short and ambiguous natural language queries are provided (e.g., ``Do you know what my favorite color is?''), which require accurate grounding in the user’s prior behavior. This setup enables evaluation of models' ability to align with diverse personal semantics. Table~\ref{tab:personabench_stats} summarizes the query and corpus statistics across users.

\begin{table}[htbp]
    \centering
    \begin{tabular}{cccccccc}
    \toprule
    \textbf{User} & \textbf{Queries} & \textbf{Corpus} & \textbf{Conv.} & \textbf{AI} & \textbf{E-com.} \\
    \midrule
    1 & 48 & 110 & 84 & 23 & 3 \\
    2 & 43 & 90  & 78 & 8  & 4 \\
    3 & 42 & 64  & 51 & 12 & 1 \\
    4 & 46 & 85  & 71 & 14 & 0 \\
    5 & 44 & 84  & 59 & 21 & 4 \\
    6 & 40 & 94  & 79 & 14 & 1 \\
    \midrule
    \textbf{Sum} & 263 & 527 & 422 & 92 & 13 \\
    \textbf{Avg} & \textbf{43.83} & \textbf{87.83} & \textbf{70.33} & \textbf{15.33} & \textbf{2.17} \\
    \bottomrule
    \end{tabular}
    \caption{Statistics of the PersonaBench dataset across six users. The \textit{Corpus} is the sum of \textit{Conv.}, \textit{AI.}, and \textit{E-com.}.}
    \label{tab:personabench_stats}
\end{table}

\paragraph{LongMemEval.}
LongMemEval~\cite{wu2025longmemeval}\footnote{\url{https://github.com/xiaowu0162/LongMemEval}} is a benchmark designed to evaluate the long-term memory capabilities of large language models, particularly their ability to retrieve task-relevant information from extended user-specific corpora. Each instance in the dataset consists of a factual question paired with a synthetic memory corpus simulating historical user data. The dataset includes two settings: \textit{LongMemEval-s}, where each question is associated with approximately 50 memories, and \textit{LongMemEval-m}, where each corpus contains over 500 memories. In total, both settings consist of 500 questions, allowing evaluation of memory retrieval precision under varying context lengths. This benchmark enables controlled analysis of personalized retrieval performance under realistic long-context scenarios.

\begin{table}[htbp]
\centering
\begin{tabular}{lcc}
\toprule
\textbf{Statistic} & \textbf{LME-s} & \textbf{LME-m} \\
\midrule
Total Questions & 500 & 500 \\
Total Memory & 25,112 & 250,948 \\
Min Memory per Question & 39 & 501 \\
Max Memory per Question & 66 & 506 \\
Avg Memory per Question & \textbf{50.22} & \textbf{501.90} \\
\bottomrule
\end{tabular}
\caption{Statistics of the LongMemEval dataset (LME). Both LME-s and LME-m contain 500 questions, with different sizes of associated memory corpora.}
\label{tab:longmemeval_stats_vertical}
\end{table}

\section*{Appendix 2: Implement Details}
\subsection*{More Detailed Implement}
\paragraph{Base Settings}:
All retrieval experiments are performed on an advanced GPU under Ubuntu 22.04.  Query expansion is generated using an advanced LLM with a temperature of 0. And each experiment we run 5 times.

\paragraph{Retrieval}: To verify the robustness of our approach, all methods are evaluated under three retrieval backbones: \texttt{multi-qa-MiniLM-L6-cos-v1}\footnote{\url{https://huggingface.co/sentence-transformers/multi-qa-MiniLM-L6-cos-v1}}, \texttt{all-MiniLM-L6-v2}\footnote{\url{https://huggingface.co/sentence-transformers/all-MiniLM-L6-v2}}, and \texttt{bge-base-en-v1.5}\footnote{\url{https://huggingface.co/BAAI/bge-base-en-v1.5}}. And FAISS to store the embedding vectors for efficient retrieval. For retrieval, we use Euclidean distance.

\subsection*{Prompt Templates for Query Expansion Methods}

\subsection*{Prompt of Baselines}

To ensure a fair comparison, we implemented the prompt-based query expansion strategies of representative baseline methods.
These prompts were carefully designed to align with the original paper. 

\begin{tcolorbox}[colback=blue!5!white, colframe=blue!75!black, title=HyDE Prompt]
Please write a paragraph that answers the question.\\
\textbf{Question:} \texttt{\{query\}}\\
\textbf{Output:}
\end{tcolorbox}

\begin{tcolorbox}[colback=blue!5!white, colframe=blue!75!black, title=Query2Term Prompt]
Answer the following question:\\
\texttt{\{query\}}\\
Give the rationale before answering.
\end{tcolorbox}

\begin{tcolorbox}[colback=blue!5!white, colframe=blue!75!black, title=MILL Prompt]
What sub-queries should be searched to answer the following query?\\
Please generate 5 sub-queries with their related passages.\\
\textbf{Question:} \texttt{\{query\}}\\
You should only return a python list like:\\
\texttt{["query1 passage1", "query2 passage2", ..., "query5 passage5"]}\\
(no comments, no markdown) without any other words and explanation.
\end{tcolorbox}

\begin{tcolorbox}[colback=blue!5!white, colframe=blue!75!black, title=Chain-of-Thought Prompt]
Solve the question step-by-step.\\
\textbf{Question:} \texttt{\{query\}}
\end{tcolorbox}

\begin{tcolorbox}[colback=blue!5!white, colframe=blue!75!black, title=ThinkQE Prompt]
Given a question \texttt{'q'} and its possible answering passages (most are wrong):\\
1. \texttt{\{d1\}}; 2. \texttt{\{d2\}}; 3. \texttt{\{d3\}}; 4. \texttt{\{d4\}}; 5. \texttt{\{d5\}}\\
Please write a correct answering passage.\\
Use your own knowledge, not just the example passages!
\end{tcolorbox}

\subsection*{Prompt Design for Our Method}


\begin{tcolorbox}[colback=blue!5!white, colframe=blue!75!black, title=P-PRF-roughly]
You are to generate 10 natural candidate utterances the user might say, inspired by the dialogue history and the current question.

\textbf{Context}
\begin{itemize}
    \item \textbf{User dialogue history (for style imitation):} \texttt{\{history\}}
    \item \textbf{Current question (to inspire the utterances):} \texttt{\{query\}}
\end{itemize}

\textbf{Guidelines}
\begin{enumerate}
    \item Generate 10 fluent, natural utterances the user might plausibly say.
    \item Do \textbf{NOT} just paraphrase; include variations in tone, emphasis, or context.
    \item Each utterance should exceed 25 words.
    \item Reflect the style and tone consistent with the document.
    \item Return ONLY valid JSON in this format (no comments, no markdown):
    
    \texttt{
    \{
    "candidates": ["...", "...", "..."]
    \}
    }
\end{enumerate}
\end{tcolorbox}

\begin{tcolorbox}[colback=blue!5!white, colframe=blue!75!black, title=P-PRF-logically]
Solve the question step-by-step, inspired by the user dialogue history.

\textbf{Context}
\begin{itemize}
    \item \textbf{User dialogue history (for style imitation):} \texttt{\{history\}}
    \item \textbf{Current question (to inspire the reasoning):} \texttt{\{query\}}
\end{itemize}

\textbf{Output:} step-by-step explanation that maintains user-specific tone and reasoning patterns.
\end{tcolorbox}

Our P-PRF involves a two-stage prompt design to jointly simulate user-specific expression  with utterance and reason. 

\section*{Appendix 3: Extend Experiment}

\subsection*{Time Efficiency}

Table~\ref{tab:time_cost} reports the latency of each stage in the ThinkQE and PBR methods, measured in seconds. Both methods share similar preprocessing overheads during index construction and user information retrieval. PBR introduces an additional graph-building step (0.0008~s) to enhance personalized query expansion, which incurs negligible computational cost. The dominant latency for both methods arises from the query expansion generation stage, accounting for approximately 2.5--2.7~s, while the final retrieval stage remains lightweight. Overall, PBR achieves personalization with marginal additional latency about 0.18~s compared to ThinkQE.

\begin{table}[h]
\centering

\begin{tabular}{lcc}
\toprule
\textbf{Stage} & \textbf{ThinkQE} & \textbf{PBR} \\
\midrule
1. Build index & 0.0797 & 0.0797 \\
2. Build graph & -- & 0.0008 \\
3. Retrieval user info & 0.0043 & 0.0043 \\
4. Call API gen QE & 2.4999 & 2.6646 \\
5. Final retrieval & 0.0064 & 0.0167 \\
\bottomrule
\end{tabular}
\caption{Time latency of each stage in ThinkQE and PBR.}
\label{tab:time_cost}
\end{table}

\subsection*{Senitivity of $k_1$}
We further conducted a hyperparameter sensitivity analysis on the LongMemEval-S dataset, focusing on the impact of $k_1$, which controls the number of top-ranked candidates used in the retrieval of the P-PRF module. As shown in Table~\ref{tab:hyper_k1}, increasing $k_1$ from 3 to 10 consistently improves both {recall} and {ndcg} across all evaluation cutoffs. The gains are particularly evident at higher recall thresholds (e.g., R@5 increases from 0.8253 to 0.8401), indicating that a larger candidate pool allows the model to capture more diverse yet relevant context signals. We set $k_1=5$ as the default in the main experiments.

\begin{table}[htbp]
\setlength\tabcolsep{2.0pt}  
\centering
\caption{Hyperparameter sensitivity on the LongMemEval-S dataset under different values of $k_1$.}
\begin{tabular}{lcccccc}
\toprule
\textbf{$k_1$} & \textbf{R@1} & \textbf{N@1} & \textbf{R@3} & \textbf{N@3} & \textbf{R@5} & \textbf{N@5} \\
\midrule
3  & 0.2076 & 0.8258 & 0.7227 & 0.8402 & 0.8253 & 0.8519 \\
5  & 0.2100 & 0.8210 & 0.7279 & 0.8444 & 0.8282 & 0.8598 \\
10 & \textbf{0.2153} & \textbf{0.8305} & \textbf{0.7470} & \textbf{0.8504} & \textbf{0.8401} & \textbf{0.8661} \\
\bottomrule
\end{tabular}
\label{tab:hyper_k1}
\end{table}

\subsection*{Generation Performance}
To further evaluate the effect of personalized query expansion on generative outcomes, we conducted an additional experiment on the LongMemEval-S dataset under \texttt{multi-qa-MiniLM-L6-cos-v1} retriever. Following prior work on RAG evaluation~\cite{adlakha2024evaluating}, we adopt a subspan Exact Match (EM) metric to account for partial correctness in generative responses. As shown in Table~\ref{tab:gen_em}, PBR outperforms baseline query expansion approaches such as HyDE and CoT. Notably, PBR achieves the highest EM score of 42.6\%, surpassing the non-personalized CoT (41.0\%) and ThinkQE (41.8\%). These results demonstrate that integrating personalized query representation not only enhances retrieval accuracy but also yields measurable gains in final generation quality, confirming the cross-stage benefit of the proposed pre-retrieval personalization.

\begin{table}[htbp]
\centering
\caption{Generation performance (EM) on the LongMemEval-S dataset.}
\begin{tabular}{lc}
\toprule
\textbf{Method} & \textbf{EM (\%)} \\
\midrule
Base & 40.0 \\
HyDE & 41.4 \\
Query2Term & 40.2 \\
MILL & 40.8 \\
CoT & 41.0 \\
ThinkQE & 41.8 \\
PBR & \textbf{42.6} \\
\bottomrule
\end{tabular}
\label{tab:gen_em}
\end{table}

\begin{table*}[ht]
\centering
\resizebox{140mm}{!}{
\begin{tabular}{lcccccccc}
\toprule
& \multicolumn{2}{c}{\textbf{PBR}} 
& \multicolumn{2}{c}{\textbf{w/o P-Anchor}} 
& \multicolumn{2}{c}{\textbf{w/o P-PRF}} 
& \multicolumn{2}{c}{\textbf{w/o both}} \\
\cmidrule(lr){2-3} \cmidrule(lr){4-5} \cmidrule(lr){6-7} \cmidrule(lr){8-9}
Metrics & R@5 & N@5 & R@5 & N@5 & R@5 & N@5 & R@5 & N@5  \\
\midrule
\rowcolor{gray!15}
\multicolumn{9}{c}{\textit{multi-qa-MiniLM-L6-cos-v1}} \\
Overall & \textbf{0.5035} & \textbf{0.4201} & 0.4871 & 0.4100 & 0.3457 & 0.2877 & 0.4484 & 0.3669 \\
Basic information & \textbf{0.5091} & \textbf{0.3647} & 0.4833 & 0.3570 & 0.3318 & 0.2242 & 0.4515 & 0.3088 \\
Preference (hard) & 0.4049 & 0.4175 & \textbf{0.4390} & \textbf{0.4295} & 0.2683 & 0.2981 & 0.3659 & 0.3759 \\
Social & \textbf{0.5541} & \textbf{0.4914} & 0.5164 & 0.4529 & 0.4277 & 0.3851 & 0.4852 & 0.4356 \\
Preference (easy) & \textbf{0.5321} & \textbf{0.5129} & 0.5192 & 0.5160 & 0.3590 & 0.3414 & 0.4904 & 0.4582 \\
\midrule
\rowcolor{gray!15}
\multicolumn{9}{c}{\textit{all-MiniLM-L6-v2}}\\
Overall & \textbf{0.4516} & \textbf{0.3855} & 0.4382 & 0.3729 & 0.2860 & 0.2449 & 0.3783 & 0.3074 \\
Basic information  & \textbf{0.4515} & \textbf{0.3485} & 0.4379 & 0.3393 & 0.1985 & 0.1617 & 0.3515 & 0.2644 \\
Preference (hard)& \textbf{0.4341} & \textbf{0.4352} & 0.4244 & 0.4201 & 0.2878 & 0.2718 & 0.4000 & 0.3547 \\
Social & \textbf{0.4494} & \textbf{0.3777} & 0.4333 & 0.3639 & 0.4179 & 0.3387 & 0.3921 & 0.3048 \\
Preference (easy)& \textbf{0.4840} & \textbf{0.4800} & 0.4712 & 0.4587 & 0.3846 & 0.3634 & 0.4295 & 0.4199 \\
\midrule
\rowcolor{gray!15}
\multicolumn{9}{c}{\textit{BAAI/bge-base-en-v1.5}} \\ 
Overall & \textbf{0.4029} & \textbf{0.3402} & 0.3921 & 0.3256 & 0.2699 & 0.2214 & 0.3738 & 0.3015 \\
Basic information & \textbf{0.4121} & \textbf{0.3057} & 0.3970 & 0.2958 & 0.2576 & 0.1804 & 0.3970 & 0.2748 \\
Preference (hard) & 0.3707 & \textbf{0.3889} & \textbf{0.3902} & 0.3778 & 0.2146 & 0.2173 & 0.3268 & 0.3343 \\
Social& \textbf{0.3657} & \textbf{0.3089} & 0.3431 & 0.2822 & 0.3261 & 0.2816 & 0.3204 & 0.2799 \\
Preference (easy) & \textbf{0.4904} & \textbf{0.4734} & 0.4744 & 0.4582 & 0.2949 & 0.2785 & 0.4583 & 0.4065 \\
\midrule
\textbf{Average} & \textbf{0.4527} & \textbf{0.3819} & 0.4391 & 0.3695 & 0.3005 & 0.2513 & 0.4002 & 0.3253 \\
\bottomrule
\end{tabular}
}
\caption{Ablation study of PBR across different retrievers and subtasks.}
\label{tab:ablation_results}
\end{table*}

\begin{table*}[ht]
\centering

\resizebox{\textwidth}{!}{
\begin{tabular}{lcccccccccccc}
\toprule
\textbf{Dataset} 
& \multicolumn{6}{c}{\textbf{LongMenEval-s}} 
& \multicolumn{6}{c}{\textbf{LongMenEval-m}} \\
Metrics  & R@1 & N@1 & R@3 & N@3 & R@5 & N@5 & R@1 & N@1 & R@3 & N@3 & R@5 & N@5 \\
\midrule
\rowcolor{gray!15}
\multicolumn{13}{c}{\textit{multi-qa-MiniLM-L6-cos-v1}} \\
PBR              & \textbf{0.2100} & \textbf{0.8210} & \textbf{0.7279} & \textbf{0.8444} & \textbf{0.8282} & \textbf{0.8598} & \textbf{0.1217} & \textbf{0.5537} & \textbf{0.4320} & \textbf{0.5776} & \textbf{0.5537} & \textbf{0.6177} \\
w/o P-Anchor       & 0.2100 & 0.8129 & 0.7199 & 0.8385 & 0.8239 & 0.8420 & 0.1169 & 0.5456 & 0.4268 & 0.5623 & 0.5480 & 0.6127 \\
w/o P-PRF          & 0.1838 & 0.7566 & 0.6277 & 0.7540 & 0.7208 & 0.7739 & 0.1122 & 0.4893 & 0.3484 & 0.4961 & 0.4368 & 0.5303 \\
w/o Both         & 0.1885 & 0.7924 & 0.6730 & 0.8020 & 0.8067 & 0.8301 & 0.1098 & 0.5251 & 0.4057 & 0.5559 & 0.5394 & 0.6038 \\
\rowcolor{gray!15}
\multicolumn{13}{c}{\textit{all-MiniLM-L6-v2}}\\
PBR              & \textbf{0.2267} & \textbf{0.8592} & \textbf{0.7780} & \textbf{0.8754} & \textbf{0.8640} & \textbf{0.8902} & \textbf{0.1408} & \textbf{0.5800} & \textbf{0.4463} & \textbf{0.5949} & \textbf{0.5847} & \textbf{0.6424} \\
w/o P-Anchor       & 0.2191 & 0.8522 & 0.7628 & 0.8622 & 0.8511 & 0.8860 & 0.1360 & 0.5795 & 0.4435 & 0.5915 & 0.5819 & 0.6356 \\
w/o P-PRF          & 0.1957 & 0.7566 & 0.6563 & 0.7765 & 0.7589 & 0.8017 & 0.1217 & 0.4797 & 0.3628 & 0.4961 & 0.4678 & 0.5415 \\
w/o Both         & 0.2196 & 0.8234 & 0.7399 & 0.8477 & 0.8568 & 0.8731 & 0.1265 & 0.5322 & 0.4415 & 0.5857 & 0.5346 & 0.6191 \\
\midrule
\rowcolor{gray!15}
\multicolumn{13}{c}{\textit{BAAI/bge-base-en-v1.5}} \\ 
PBR              & \textbf{0.2315} & \textbf{0.8902} & \textbf{0.8138} & \textbf{0.9072} & \textbf{0.9045} & \textbf{0.9191} & \textbf{0.1408} & \textbf{0.6205} & \textbf{0.5489} & \textbf{0.6612} & \textbf{0.6874} & \textbf{0.7060} \\
w/o P-Anchor       & 0.2239 & 0.8850 & 0.8121 & 0.9037 & 0.8993 & 0.9154 & 0.1304 & 0.6125 & 0.5322 & 0.6579 & 0.6721 & 0.7007 \\
w/o P-PRF          & 0.2029 & 0.8234 & 0.7446 & 0.8517 & 0.8449 & 0.8704 & 0.1002 & 0.5298 & 0.4344 & 0.5698 & 0.5680 & 0.6125 \\
w/o Both         & 0.2267 & 0.8616 & 0.7900 & 0.8863 & 0.8926 & 0.9007 & 0.1217 & 0.5943 & 0.5322 & 0.6605 & 0.6444 & 0.6929 \\
\bottomrule
\end{tabular}
}
\caption{Ablation results of PBR on LongMenEval-s and LongMenEval-m datasets.}
\label{tab:ablation_longmen}
\end{table*}

\subsection*{Ablation Study in Two Datasets}
\paragraph{Complementary Ablation Insights.}  
Across both the retrieval backbones and datasets, we observe a \textit{consistent performance drop when either component (P-Anchor or P-PRF) is removed}, confirming their complementary contributions to the overall effectiveness of the PBR framework.

First, \textbf{removing P-PRF leads to the most significant degradation}, especially on more complex queries such as \texttt{Preference (hard)} and datasets like LongMenEval-m, where user intent is more nuanced. For example, on \texttt{multi-qa-MiniLM} in Table~\ref{tab:ablation_results}, R@5 drops from 0.5035 to 0.3457 without P-PRF, and similarly in Table~\ref{tab:ablation_longmen}, LongMenEval-s sees a drop in R@3 from 0.7780 to 0.6563 without P-PRF. This highlights the \textit{critical role of user-aware pseudo relevance feedback in capturing latent intent}.

Second, \textbf{removing P-Anchor has a more moderate but consistent impact}, especially in tasks requiring robust grounding of queries within diverse semantic corpora (e.g., \texttt{Social} and \texttt{Preference (easy)}). This suggests that \textit{style-consistent anchoring helps stabilize query representations}, particularly when user content is lexically varied.

Finally, the combination of removing both components consistently results in a \textbf{further drop}, but not always the worst performance. This indicates a certain degree of \textit{redundancy or shared effect between P-Anchor and P-PRF}, suggesting that while they operate on different principles (style anchoring vs. semantic feedback), they jointly reinforce personalized query understanding. On average (last row, Table~\ref{tab:ablation_results}), the full PBR outperforms the no-component baseline by \textbf{+13.1\% R@5} and \textbf{+17.4\% N@5}.

\subsection*{Effectiveness of P-PRF in Two Datasets}

\begin{figure*}[ht]
    \centering
    \includegraphics[width=0.98\linewidth]{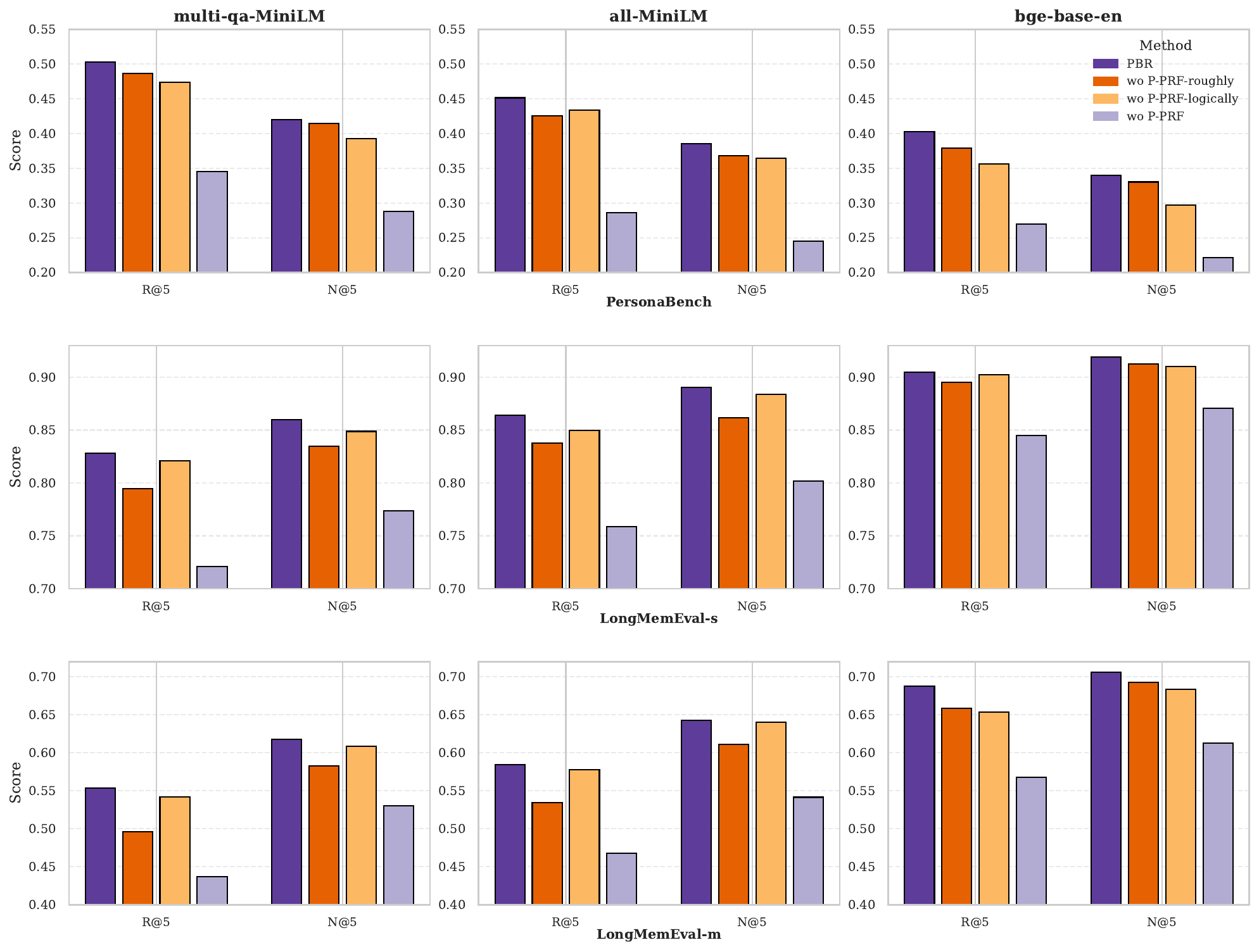}
    \caption{%
        Overall performance in \textit{Personabench} comparison across three ablation retrieval modes. 
    }
    \label{fig:pprf-ablation-all}
    \vspace{-10pt}
\end{figure*}

\paragraph{Visualization-based Analysis.} 
Figure~\ref{fig:pprf-ablation-all} presents a comparative bar plot of PBR and its ablated variants across three retrievers (multi-qa-MiniLM, all-MiniLM, bge-base-en) and three evaluation datasets (PersonaBench, LongMenEval-s, LongMenEval-m). Each subplot reports the R@5 and N@5 metrics, offering insight into both retrieval recall and ranking quality.

\textbf{(1) On \textit{PersonaBench}}, PBR consistently outperforms all variants across all retrievers. Notably, the drop in N@5 when removing P-PRF is more severe than the drop in R@5, indicating that pseudo-relevance feedback contributes more to fine-grained ranking alignment than coarse recall.

\textbf{(2) On \textit{LongMenEval-s}}, where queries are more succinct, the gap between PBR and its variants becomes more pronounced, especially under bge-base-en. This suggests that PRF-based refinements are crucial for handling underspecified queries by injecting user-grounded semantics into the expansion.

\textbf{(3) On \textit{LongMenEval-m}}, the performance degradation from removing P-PRF is still significant but relatively less than in LongMenEval-s. This is likely due to longer query inputs providing stronger initial signals, slightly compensating for the lack of PRF. However, the logical and rough removal of PRF still lags behind full PBR, validating the importance of structurally grounded expansion paths.

Overall, the visual analysis confirms the complementary roles of anchor-based grounding and pseudo-feedback refinement in boosting personalized retrieval quality across varied retriever and dataset settings.

\end{document}